\documentclass[aps,prd,preprint,groupedaddress,showpacs,A4paper]{revtex4-1}

\usepackage{graphicx}						
\usepackage{amsmath,amsfonts,amsbsy}
\usepackage{slashed}

\usepackage{tikz}

\usetikzlibrary{intersections}
\usetikzlibrary{decorations.pathmorphing}
\usetikzlibrary{decorations.markings}

\tikzset{
    laser/.style={decorate, decoration={snake,segment length=5mm}, draw=black},
    electron/.style={draw=black, postaction={decorate},
        decoration={markings,mark=at position .55 with {\arrow[draw=black]{>}}}}, 
}

\newcommand{\psl}{\slashed{p}}
\newcommand{\asl}{\slashed{a}}
\newcommand{\Dsl}{\slashed{D}}

\newcommand{\Asl}{\slashed{A}}
\newcommand{\ksl}{\slashed{k}}
\newcommand{\msl}{\!\slashed{\,m}}
\newcommand{\mslb}{\slashed{\bar{{m}}}}
\newcommand{\mslt}{\slashed{\tilde{m}}}

\newcommand{\pbar}{{\bar{p}}}
\newcommand{\qbar}{{\bar{q}}}
\newcommand{\pbsl}{\slashed{\pbar}}

\newcommand{\U}{\mathcal{U}}
\newcommand{\V}{\mathcal{V}}
\newcommand{\Uv}{\mathcal{U}_{_{\mathrm{V}}}}

\newcommand{\Vv}{\mathcal{V}_{_{\mathrm{V}}}}
\newcommand{\Uvb}{\bar{\mathcal{U}}_{_{\mathrm{V}}}}

\newcommand{\Vvb}{\bar{\mathcal{V}}_{_{\mathrm{V}}}}
\newcommand{\Fcal}{\mathcal{F}_\ell}
\newcommand{\Jcal}{\mathcal{J}}

\newcommand{\psiV}{\psi_{_{\mathrm{V}}}}
\newcommand{\psibV}{\bar{\psi}_{_{\mathrm{V}}}}

\newcommand{\aVa}{a^{(\alpha)}_{_{\mathrm{V}}}\!}

\newcommand{\bVb}{b^{(\beta)}_{_{\mathrm{V}}}\!}

\newcommand{\adVb}{a^{\dag(\beta)}_{_{\mathrm{V}}}\!}
\newcommand{\bdVa}{b^{\dag(\alpha)}_{_{\mathrm{V}}}\!}

\newcommand{\Jdn}{\Jcal_n^\Delta}

\newcommand{\Dl}{D_{_{\mathrm{\ell}}}}
\newcommand{\Dld}{D^{\dag}_{_{\mathrm{\ell}}}}
\newcommand{\intfps}{\int{\bar{}\kern-0.45em d}^{\,3}p\,}
\newcommand{\intps}{\int\frac{{\bar{}\kern-0.45em d}^{\,3}p\,}{2E^*_p}}
\newcommand{\intfp}{\int{\bar{}\kern-0.45em d}^{\,4}p\,}
\newcommand{\intfqs}{\int{\bar{}\kern-0.45em d}^{\,3}q\,}
\newcommand{\intfpqs}{\int{\bar{}\kern-0.45em d}^{\,3}p\,\,\,{\bar{}\kern-0.45em d}^{\,3}q\,}
\newcommand{\bra}[1]{\langle #1|}
\newcommand{\ket}[1]{|#1\rangle}
\newcommand{\braV}[1]{{}_{_{\mathrm{V}}}\!\langle #1|}
\newcommand{\ketV}[1]{|#1\rangle_{_{\mathrm{V}}}}

\begin{document}

\title{The Fermionic Propagator in an Intense Background}

\author{Martin~Lavelle and David~McMullan}
\affiliation{Centre for Mathematical Sciences\\Plymouth  University\\
Plymouth, PL4 8AA, UK}


\begin{abstract}
New results for the fermion propagator in a laser background are presented. We show that the all orders electron propagator can be written in a compact and appealing form as a sum of sideband poles with a matrix wave function renormalisation and a matrix valued mass shift. This last result is essential in the fermionic theory if we are to maintain that both the mass and its square pick up a correction only at  order $e^2$. A perturbative verification of our results is carried out.
\end{abstract}



\maketitle

\section{Introduction}
In this paper we will discuss fermionic propagation in background fields. Specifically we are interested in electrons propagating in a plane wave background~\cite{Volkov:1935zz}. This is important for understanding  recent work with high power laser facilities~\cite{Heinzl:2011ur}\cite{DiPiazza:2011tq}. The result we will obtain is simple and has a clear physical interpretation. A surprise is that all the renormalisation factors are matrices. 

The propagator is one of the basic building blocks linking physical processes with the structure of quantum field theory. We recall that for a scalar field, $\phi(x)$, the Feynman propagator is given by the time ordered product
\begin{equation}
	\bra0 \mathrm{T} \phi(x)\phi^{\dagger}(y)\ket0 =
	\intfp e^{-ip{\cdot}(x-y)}\frac i{p^2-m^2+i\epsilon}\,,
\end{equation}
where the fields are taken to be free and the bar in the measure indicates appropriate factors of $2\pi$. In the presence of loop corrections the form of the propagator is preserved and typically becomes  
\begin{equation}
	\bra0 \mathrm{T} \phi(x)\phi^{\dagger}(y)\ket0 =
	\intfp e^{-ip{\cdot}(x-y)}\frac {iZ_2}{p^2-(m^2+\delta m^2)+i\epsilon}\,,
\end{equation}
which can be interpreted as the propagation of renormalised fields with a renormalised mass. 

A similar, but richer, structure in the propagator is  found when the scalar field interacts with a laser plane wave background~\cite{Reiss:1966A}\cite{Brown:1964zz}\cite{Neville:1971uc}\cite{Dittrich:1973rn}\cite{Dittrich:1973rm}\cite{Kibble:1975vz}\cite{Mitter:1974yg}\cite{Ritus:1985review}\cite{Ilderton:2012qe}\cite{Lavelle:2013wx}\cite{Lavelle:2014mka} 
\begin{equation}
 \sum_{n=-\infty}^\infty \intfp e^{-ip{\cdot}(x-y)}\frac{iZ_2^{(n)}}{(p+nk)^2-(m^2+\delta m^2)+i\epsilon}\,,
\end{equation}
where we see a sum over so-called sideband states~\cite{Reiss:1962}\cite{Reiss:2009ed}\cite{Ritus:1972ky} and note that the mass shift is common to all sidebands. The exact form of the mass and the momentum dependent wave function renormalisation factors depend on the details of the laser background. For a linearly polarised background, the propagator has the form 
\begin{equation}\label{eq:scalarprop}
 \sum_{n=-\infty}^\infty \intfp e^{-ip{\cdot}(x-y)}\frac{i\mathrm{J}^2_n(eu_p,e^2v_p)}{(p+nk)^2-m_*^{2}+i\epsilon}\,,
\end{equation}
where $u_p$ and $v_p$ are laser parameters that we define in~(\ref{eq:uvdefs}). The mass shift is here given by $m^2_*=m^2-\frac12e^2 a^2$ where $a_\mu$ is the (spacelike) laser amplitude and this mass shift does not acquire higher order corrections. The wave function renormalisation for the various sidebands is written  in terms of a product of generalised Bessel functions which can be expressed in terms of normal Bessel functions as
\begin{equation}\label{bes9}
    \mathrm{J}_n(eu_p,e^2v_p)=\sum_{r=-\infty}^\infty J_{n-2r}(eu_p)J_r(e^2v_p)\,.
\end{equation}
The fine details in (\ref{eq:scalarprop}) are polarisation dependent but the broad structures for a plane wave laser background, including the sum over sidebands and the mass shift being at order $e^2$ only, are common. 
 
Note that in the literature, the two point function in a laser background is widely referred to as the propagator. However, through interactions with the background, this two point function includes diagrams where the initial and final momenta of the matter field are distinct. In this paper, and in our summary above, the focus is on the diagonal part of the two point function which does not include momentum altering vertex  effects. This is what we call the propagator in this paper and this is given for the scalar theory in~(\ref{eq:scalarprop}).

As in our previous work, we are guided below by experience from infra-red physics. A   charge propagating in a laser  is  indistinguishable from a charge which absorbs and emits the same number of laser photons. These laser induced degeneracies  parallel  the soft and collinear degeneracies of the infra-red regime in  QED and QCD~\cite{Bloch:1937pw}\cite{Kinoshita:1962ur}\cite{Lee:1964is}\cite{Lavelle:2005bt}. Understanding the  mass shift in a laser background may also help to clarify the current versus constituent mass distinction in QCD~\cite{Lavelle:1995ty}.  

Turning now to fermionic QED, the perturbative propagator without a background has the form
\begin{align}
\begin{split}
	\bra0 \mathrm{T} \psi(x)\bar\psi(y)\ket0 = &
\intfp e^{-ip{\cdot}(x-y)}\frac {iZ_2}{\psl-(m+\delta m)+i\epsilon}\,,
	\\
	=& \intfp e^{-ip{\cdot}(x-y)}\frac {iZ_2(\psl+m+\delta m)}{p^2-(m+\delta m)^2+i\epsilon}\,,\label{eq:fermifreeish}
	\end{split}
\end{align}
where we note the presence of single powers of $m$.

The extension of this theory to plane wave laser backgrounds  has been previously considered~\cite{Brown:1964zz,  Reiss:1966A,Becker:1979rc,Bergou:1980xh}. The results obtained also display the sideband structure of the scalar theory but the two point function is quite involved, see, e.g., (5.7) and (5.8) in Ref.~\cite{Reiss:1966A}, and does not clearly resemble a renormalised propagator. In particular in the numerator of the propagator obtained by Reiss and Eberly~\cite{Reiss:1966A} there are two gamma matrix structures, one proportional to $\psl-e\Asl+m $ and one proportional to the laser momentum, $\ksl$, each multiplied by combinations of Bessel functions. Similarly, additional gamma matrix structures are obtained in Appendix~A of Brown and Kibble~\cite{Brown:1964zz} and by Ritus~\cite{Ritus:1985review}.

The mass shift in the fermionic theory also satisfies the spin independent relation~\cite{Brown:1964zz,Reiss:1966A}  
\begin{equation}\label{eq:massy}
	m^2_*=(m+\delta m)^2= m^2-\tfrac12e^2 a^2\,,
\end{equation}
to all orders in the coupling. This might naturally suggest, see for example section~40 of~\cite{Berestetsky:1982aq} and more recently~\cite{Meuren:2011hv}, that 
\begin{equation}
	m_*=m+\delta m= m\sqrt{1-\frac{e^2a^2}{2m^2}}\,,
\end{equation}
which would, however, mean that $\delta m$ would acquire corrections to all orders in the coupling through an expansion of the above.

Below we will first show that an explicit perturbative calculation disagrees with this expectation for the mass shift. We will then present an all orders calculation of the fermion propagator which will reveal a simple form which can be interpreted, as for the scalar theory, in terms of mass and wave function renormalisation factors. An unexpected feature is that all the renormalisation factors we find, including the mass shift, are matrices. The mass shift will itself only be corrected at order $e^2$, but still satisfy~(\ref{eq:massy}) to all orders. As this is  surprising we will independently verify all of the  structures we obtain via a low order perturbative calculation. Finally we will present some conclusions. Some technical details are presented in appendices.

\section{Perturbative expansion} 
\label{sec:pert}

The Lagrangian density describing the interactions of a fermion with the laser is 
\begin{equation}\label{eq:Laggy}
	{\cal{L}}=\bar\psi(i\Dsl-m)\psi\,.
\end{equation}
In this the covariant derivative is given by $D_\mu=\partial_\mu+ieA_\mu$ and, in a linearly polarised background, 
\begin{equation}
	A_\mu(x)=a_\mu\cos(k\cdot x)\,,
\end{equation}
where $k=k_0(1,0,0,1)$ is the null vector characterising the laser. Note that the gauge fixing condition used is $k\cdot a=0$.

 The Feynman rules for the vertices in this linearly polarised theory are given in Fig.~1. For this polarisation, the rules are identical for the absorption and the emission of a laser photon.
\begin{figure}\label{rules}
\begin{center}
\begin{minipage}[c]{0.6\linewidth}
\begin{tikzpicture}[xscale=0.6, yscale=0.75]
 \draw[electron] (0,0) -- node[below=4.5pt]{$p$}(2,0);
 \draw[laser] (0,1.5) -- (2,0);
 \draw[electron] (2,0) -- node[below=2pt]{$p+k$}(4,0);
 \fill (2, 0) circle (1.5pt);
 \draw[white] (4,0) -- node [right=2pt]
{$\displaystyle\color{black}=
$}(4,1.5);
\end{tikzpicture}
\begin{tikzpicture}[xscale=0.6, yscale=0.75]
 \draw[electron] (0,0) -- node[below=4.5pt]{$p$}(2,0);
 \draw[laser] (4,1.5) -- (2,0);
 \draw[electron] (2,0) -- node[below=2pt]{$p-k$}(4,0);
 \fill (2, 0) circle (1.5pt);
  \draw[white] (4,0) -- node [right=2pt]
{$\displaystyle\color{black}=
\frac{ie}2 \asl $}(4,1.5);
\end{tikzpicture}
\end{minipage}
\caption{\label{lin3Frules261012}The three point Feynman rules for linear polarisation}
\end{center}
\end{figure}
The two diagrams that contribute at lowest order in the coupling are shown in Fig.~2.
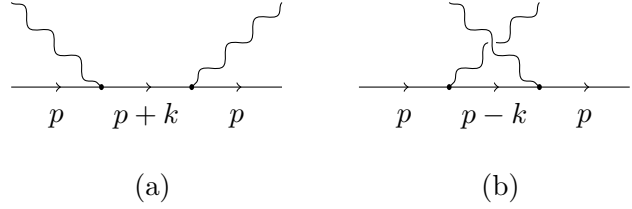
\begin{figure}\label{lowfigs}
\begin{center}
\begin{minipage}[c]{0.25\linewidth}
\begin{tikzpicture}[xscale=0.6, yscale=0.75]
 \draw[electron] (0,0) -- node[below=4.5pt]{$p$}(2,0);
 \draw[laser] (0,1.5) -- (2,0);
 \draw[laser](6,1.5) -- (4,0);
 \draw[electron] (2,0) -- node[below=2pt]{$p+k$}(4,0);
 \draw[electron] (4,0) -- node[below=4.5pt]{$p$}(6,0);
 \fill (2, 0) circle (1.5pt);
 \fill (4,0) circle (1.5pt);
\end{tikzpicture}
\newline
\centerline{(a)}
\end{minipage}
\quad
\begin{minipage}[c]{0.25\linewidth}
\begin{tikzpicture}[xscale=0.6, yscale=0.75]
 \draw[electron] (0,0) -- node[below=4.5pt]{$p$}(2,0);
 \draw[laser](4,1.5) -- (2,0);
 \fill[white] (2.95,0.78) circle (2.5pt);
\draw[laser] (2,1.5) -- (4,0);
 \draw[electron] (2,0) -- node[below=2pt]{$p-k$}(4,0);
 \draw[electron] (4,0) -- node[below=4.5pt]{$p$}(6,0);
 \fill (2, 0) circle (1.5pt);
 \fill (4,0) circle (1.5pt);
\end{tikzpicture}
\newline
\centerline{(b)}
\end{minipage}
\end{center}
\caption{\label{line2propdiags} Leading order contributions to the propagator}
\end{figure}
The first diagram yields after a partial fraction expansion
\begin{equation}
	\frac{ie^2}{16(p{\cdot} k)^2} (\psl+m)\asl(\psl+\ksl+m)\asl(\psl+m)
\left(
-\frac1{p^2-m^2}+\frac{2p{\cdot} k}{(p^2-m^2)^2}+\frac{1}{(p+k)^2-m^2}
\right)\,,
\end{equation}
and the second diagram is obtained by replacing $k\to-k$. In our perturbative calculations we suppress the $i\epsilon$ prescription. The gamma matrix structures are now expanded depending upon whether the denominator contains poles in $p^2-m^2$ or is in an adjacent sideband. In this initial discussion, we shall restrict ourselves to the central sideband alone. In the double pole coefficient the $\ksl$~term alone survives when the two diagrams are added, while in the single pole coefficient the $\ksl$~term cancels. Thus the two diagrams contribute to the central sideband
\begin{align}\label{eq:addedcentral}
	\frac{ie^2}{8(p{\cdot} k)^2} \left(-(\psl+m)\asl(\psl+m)\asl(\psl+m)
\frac1{p^2-m^2}
+
(\psl+m)\asl\ksl\asl(\psl+m)\frac{2p{\cdot} k}{(p^2-m^2)^2}
\right)\,.
\end{align}
In the first term we rewrite
\begin{equation}\label{eq:gammatrick}
	(\psl+m)\asl(\psl+m)\asl(\psl+m)=
(\psl+m)\asl\left(
-(p^2-m^2)\asl+2p{\cdot} a (\psl+m)
\right)
\,.
\end{equation}
The first part here will not contribute any pole to the central sideband and we have verified that the sum of all such terms without poles cancel when the contributions from all sidebands are summed. We will therefore just drop them below. The double pole term in~(\ref{eq:addedcentral}) will also generate single pole terms in a similar fashion. The sum of all the terms containing poles in the central sideband is found to be 
\begin{align}\label{eq:pertcentraltwo}
	{ie^2} \left(-\frac{(p{\cdot} a)^2}{2(p{\cdot} k)^2}\frac{\psl+m}{p^2-m^2}
+\frac{ a^2\ksl }{4p{\cdot} k}\frac{1}{p^2-m^2}
-\frac12a^2\frac{\psl+m}{(p^2-m^2)^2}
\right)\,.
\end{align}
We should now compare this with the general form of~(\ref{eq:fermifreeish}). At this leading order we expect three contributions. Writing $Z_2^{(0)}=1+\delta Z_2^{(0)}$ (where $\delta Z_2^{(0)}$  is expected to have a matrix form) this would give
\begin{equation}\label{eq:naiveexpectations}
	ie^2\left(
\frac{\delta Z_2^{(0)}(\psl+m)}{p^2-m^2}
+\delta m\frac{1}{p^2-m^2}
-\frac12a^2 \frac{1}{(p^2-m^2)^2}
	\right)\,,
\end{equation}
where we have used~(\ref{eq:massy}) in the last term.

In~(\ref{eq:pertcentraltwo}) the double pole term is the expected mass shift from the expansion of the denominator of the central sideband. The first term gives us information about the wave function renormalisation factor. However, the second term in~(\ref{eq:pertcentraltwo}), which involves  $a^2$ as would be expected of a mass shift, contains a $\ksl$-factor and is not obtained by simply expanding the numerator in~(\ref{eq:fermifreeish}). We will therefore now carry out an all orders calculation of the propagator to resolve this puzzle and also determine the form of $Z_2^{(n)}$.

\section{Volkov fermion} 
\label{sec:volkov_fermion}

We now want to construct the spinor field $\psiV(x)$ which satisfies 
\begin{equation}\label{eq:eqm}
	(i\Dsl-m)\psiV(x)=0\,.
\end{equation}
 This Volkov equation (\ref{eq:eqm}) is   solved by 
\begin{align}\label{eq:fvolkov}
\begin{split}
	\psiV(x)=\intfps\frac{m}{E^*_p}&\Bigg(\Fcal(x,\pbar)\Dl(x,\pbar)\Uv^{(\alpha)}(p)\aVa(p)\\&\qquad+\Fcal(x,-\pbar)\Dl(x,-\pbar)\Vv^{(\alpha)}(p)\bdVa(p)\Bigg)\,.	
\end{split}	
\end{align}
There is a lot of structure to this solution that we need to unpack carefully. 

The on-shell momentum $\pbar_\mu=(E^*_p,\underline{p})$ satisfies the shifted on-shell condition of the quasi-momentum:
\begin{equation}
	\pbar^2-m^2_*=E_p^{*\,2}-\underline{p}^{\,2}-m^2_*=0\,.
\end{equation}

Just as for the scalar case,  plane waves are distorted in this background and become equal to
\begin{align}\label{eq:Ddefinition}
\begin{split}
  \Dl(x,\pbar)&=e^{-i\pbar{\cdot}x}e^{i(eu_\pbar\sin(k\cdot x)+e^2v_\pbar\sin(2k\cdot x))}\,, \\
  & =e^{-i\pbar{\cdot}x}\sum_{n=-\infty}^\infty e^{in k \cdot x}{\mathrm{J}}_n(eu_\pbar,e^2v_\pbar)\,.
 \end{split}
\end{align}
In this expression we have defined the laser variables $u_\pbar$ and $v_\pbar$ by
\begin{equation}\label{eq:uvdefs}
 	u_\pbar=-\frac{\pbar{\cdot} a}{\pbar{\cdot} k}\quad \mathrm{and} \quad v_\pbar=\frac{a^2}{8\pbar{\cdot} k}\,.
 \end{equation}

The rest of equation~(\ref{eq:fvolkov})  reflects the fermionic nature of QED. Focusing initially on the first term in $\psiV(x)$, we see that in the Volkov equation the derivative will only act on the factors $\Fcal(x,\pbar)\Dl(x,\pbar)$. The prefactor $\Fcal(x,\pbar)$ is a matrix given by
\begin{equation}\label{eq:prevol1}
	\Fcal(x,\pbar)=\left(1+e\frac{\ksl\Asl(x)}{2\pbar\cdot k}\right)\,.
\end{equation}
The action of the covariant derivative is then given by
\begin{equation}\label{eq:prevol2}
	i\Dsl\Fcal(x,\pbar)\Dl(x,\pbar)=\Fcal(x,\pbar)\Dl(x,\pbar)(\pbsl+2e^2v_\pbar\ksl)\,.
\end{equation}
Similarly, in the second part of~(\ref{eq:fvolkov}) we get
\begin{equation}
	i\Dsl\Fcal(x,-\pbar)\Dl(x,-\pbar)=\Fcal(x,-\pbar)\Dl(x,-\pbar)(-\pbsl-2e^2v_\pbar\ksl)\,.
\end{equation}
The Volkov spinors are then introduced to satisfy 
\begin{equation}\label{eq:uvolkov}
	(\pbsl+2e^2v_\pbar\ksl)\Uv^{(\alpha)}(p)=m\,\Uv^{(\alpha)}(p)\,,
\end{equation}
and
\begin{equation}\label{eq:vvolkov}
	(\pbsl+2e^2v_\pbar\ksl)\Vv^{(\alpha)}(p)=-m\,\Vv^{(\alpha)}(p)\,.
\end{equation}
The precise form for these Volkov spinors is presented in Appendix~A.

Combining these results we see that $\psiV(x)$ as defined by equation~(\ref{eq:fvolkov}) does indeed satisfy the Volkov equation~(\ref{eq:eqm}).

Having constructed the Volkov field, $\psiV(x)$, we now simply state the expression for the Dirac adjoint  
\begin{align}\label{eq:fbarvolkov}
\begin{split}
	\psibV(y)=\intfqs\frac{m}{E^*_q}&\Bigg(\Uvb^{(\beta)}(q)\Fcal(y,-\qbar)\Dld(y,\qbar)\adVb(q)\\&\qquad+\Vvb^{(\beta)}(q)\Fcal(y,\qbar)\Dld(y,-\qbar)\bVb(q)\Bigg)\,.	
\end{split}
\end{align}     

Returning to the on-shell condition (\ref{eq:uvolkov}) for the Volkov spinors, we can directly reinterpret this as
\begin{equation}\label{eq:uvolkov1}
	\pbsl\,\Uv^{(\alpha)}(p)=(m-2e^2v_\pbar\ksl)\,\Uv^{(\alpha)}(p)\,,
\end{equation}
that is, in the spinor theory, the mass shift has a matrix form with $m_*=m+\delta \mslb$ with
\begin{equation}\label{eq:massmatrix}
	\delta \mslb =-2e^2v_\pbar\ksl=-\frac{e^2a^2}{4\pbar{\cdot}k}\ksl\,.
\end{equation}
 Let us now show that this is consistent with the result familiar from the scalar theory. We multiply both sides of (\ref{eq:uvolkov1}) by $\pbsl$ and anticommute the $\pbsl$ factor through the $\ksl$ to obtain 
\begin{equation}\label{eq:uvolkov2}
	\pbar^2\,\Uv^{(\alpha)}(p)=(m\pbsl-4e^2v_\pbar\pbar\cdot k +2e^2v_\pbar\ksl\pbsl)\,\Uv^{(\alpha)}(p)\,.
\end{equation}
We now reuse (\ref{eq:uvolkov1}) in the final term and, recalling that $k^2=0$, obtain the spin-independent on-shell result 
\begin{equation}
	\pbar^2\,\Uv^{(\alpha)}(p)=m^2_*\,\Uv^{(\alpha)}(p)=(m^2-\tfrac12e^2 a^2)\,\Uv^{(\alpha)}(p)\,.
\end{equation}
 Note that this requires $\delta \mslb$ to be a matrix, as one would otherwise  generate corrections to $m^2$ at order $e^4$. 

 However, we cannot yet identify $\delta m$ in~(\ref{eq:naiveexpectations}) from this, as if we were to replace $\delta m$ by $\delta \mslb$  we would obtain~(\ref{eq:pertcentraltwo}) but with the wrong sign for the $\ksl$~term. We shall though see later that the result for  $\delta \mslb$ is correct and that the naive expectation~(\ref{eq:naiveexpectations}) is wrong.


\section{The propagator} 
\label{sec:the_propagator}
We are now in a position to  calculate the Volkov propagator
\begin{align}
\begin{split}
	\braV{0}\mathrm{T}\psiV(x)\psibV(y)\ketV{0}=\theta(x^0-y^0)\,\braV{0}&\psiV(x)\psibV(y)\ketV{0}\\&-\theta(y^0-x^0)\,\braV{0}\psibV(y)\psiV(x)\ketV{0}\,,
\end{split}
\end{align}
where the Volkov vacuum is identified through the modes in~(\ref{eq:fvolkov}) and~(\ref{eq:fbarvolkov}). Note that the creation and annihilation operators satisfy 
\begin{equation}
	\{ \aVa(p),\adVb(q)\}
	=
	(2\pi)^3\frac{E_p^*}m  \delta^{\alpha\beta} 
	\delta^{(3)}(p-q)\,.
\end{equation}
Looking at the first vacuum expectation value we have
\begin{equation}\label{eq:vac1}
	\braV{0}\psiV(x)\psibV(y)\ket{0}=\intfps\frac{1}{2E^{*}_p}\Fcal(x,\pbar)(\pbsl+m- \delta\mslb)\Fcal(y,-\pbar)\Dl(x,\pbar)\Dld(y,\pbar)\,.
\end{equation} 
In a similar way  we obtain
\begin{equation}\label{eq:vac2}
	\braV{0}\psibV(y)\psiV(x)\ketV{0}=\intfps\frac{1}{2E^{*}_p}\Fcal(x,-\pbar)(\pbsl-m- \delta\mslb)\Fcal(y,\pbar)\Dl(x,-\pbar)\Dld(y,-\pbar)\,.
\end{equation}
We now use these to build up the propagator. To this end we require the following results which follow from~(\ref{eq:Ddefinition}) 
\begin{equation}
	\Fcal(x,\pbar)\Dl(x,\pbar)=\sum_{n=-\infty}^{\infty} e^{-i\pbar{\cdot}x}e^{ink{\cdot x}}\Big(\Jcal_n(\pbar)+e\frac{\ksl\asl}{4\pbar{\cdot}k}\Jdn(\pbar)\Big)\,,
\end{equation}
where we have condensed our notation for the generalised Bessel function so that
\begin{equation}
	\Jcal_n(\pbar)=\mathrm{J}_n(eu_{\pbar},e^2v_{\pbar})\,,
\end{equation}
and
\begin{equation}
	\Jdn(\pbar)=\Jcal_{n-1}(\pbar)+\Jcal_{n+1}(\pbar)\,.
\end{equation}
In a similar way we obtain
\begin{equation}
	\Fcal(x,-\pbar)\Dl(x,-\pbar)=\sum_{n=-\infty}^{\infty} e^{i\pbar{\cdot}x}e^{ink{\cdot}x}\Big(\Jcal_n(-\pbar)+e\frac{\asl\ksl}{4\pbar{\cdot}k}\Jdn(-\pbar)\Big)\,,
\end{equation}
while
\begin{equation}
	\Fcal(y,-\pbar)\Dld(y,\pbar)=\sum_{n=-\infty}^{\infty} e^{i\pbar{\cdot}y}e^{-ink{\cdot}y}\Big(\Jcal_n(\pbar)+e\frac{\asl\ksl}{4\pbar{\cdot}k}\Jdn(\pbar)\Big)\,,
\end{equation}
and
\begin{equation}
	\Fcal(y,\pbar)\Dld(y,-\pbar)=\sum_{n=-\infty}^{\infty} e^{-i\pbar{\cdot}y}e^{-ink{\cdot}y}\Big(\Jcal_n(-\pbar)+e\frac{\ksl\asl}{4\pbar{\cdot}k}\Jdn(-\pbar)\Big)\,.
\end{equation}
Inserting these expansions into (\ref{eq:vac1}) and (\ref{eq:vac2}) and using the decomposition
\begin{equation}
	\psiV(x)=\sum_{n=-\infty}^\infty \psi_n(x)\,,
\end{equation}
gives  the propagator defined through the diagonal contributions to the two point function
\begin{align}
\begin{split}
 	\braV{0}\psi_n(x)\bar{\psi}_n(y)\ketV{0}&=\intfps \frac{1}{2E^{*}_p}e^{-i\pbar{\cdot}(x-y)}e^{ink{\cdot}(x-y)}\Big(\Jcal_n(\pbar)+e\frac{\ksl\asl}{4\pbar{\cdot}k}\Jdn(\pbar)\Big)\\&\qquad\times(\pbsl+m- \delta\mslb	)\Big(\Jcal_n(\pbar)+e\frac{\asl\ksl}{4\pbar{\cdot}k}\Jdn(\pbar)\Big)\,,
 \end{split}
 \end{align}
 and
 \begin{align}
 \begin{split}
 	\braV{0}\bar{\psi}_n(y)\psi_n(x)\ketV{0}&=\intfps \frac{1}{2E^{*}_p}e^{i\pbar{\cdot}(x-y)}e^{ink{\cdot}(x-y)}\Big(\Jcal_n(-\pbar)+e\frac{\asl\ksl}{4\pbar{\cdot}k}\Jdn(-\pbar)\Big)\\&\qquad\times(\pbsl-m- \delta\mslb)\Big(\Jcal_n(-\pbar)+e\frac{\ksl\asl}{4\pbar{\cdot}k}\Jdn(-\pbar)\Big)\,.
 	\end{split}
 \end{align}
 The Volkov propagator then becomes equal to the sum over $n$ of
\begin{align}\label{eq:vstep1}
\begin{split}
-\int\!\!\frac{dp^0}{2 \pi}\frac{{\bar{}\kern-0.45em d}^{\,3}p}{2E^*_p}&\frac{i}{p^0-i \epsilon}
\Big(
e^{i(p^0-E^*_p+nk^0)(x^0-y^0)}e^{i(\underline{p}-n\underline{k})\cdot(\underline{x}-\underline{y})}\\
&\times \big(\Jcal_n(\pbar)+e\frac{\ksl\asl}{4\pbar{\cdot} k}\Jdn(\pbar)\big)(\pbsl+m- \delta\mslb)\big(\Jcal_n(\pbar)+e\frac{\asl\ksl}{4\pbar{\cdot} k}\Jdn(\pbar)\big)\\
&\qquad -e^{-i(p^0-E^*_p-nk^0)(x^0-y^0)}e^{-i(\underline{p}+n\underline{k})\cdot(\underline{x}-\underline{y})}\\
&\times \big(\Jcal_n(-\pbar)+e\frac{\asl\ksl}{4\pbar{\cdot} k}\Jdn(-\pbar)\big)(\pbsl-m- \delta\mslb)\big(\Jcal_n(-\pbar)+e\frac{\ksl\asl}{4\pbar{\cdot} k}\Jdn(-\pbar)\big)
\Big)\,.
\end{split}
\end{align}
Combining the variables $p^0$ and $p^i$ into an off shell four moment $p^\mu$ (details of which are given in Appendix~B) we can write the fermionic Volkov  propagator as
 \begin{align}\label{eq:final1}
 \begin{split}
 	\sum_{n=-\infty}^\infty\int{\bar{}\kern-0.45em d}^{\,4}p \,e^{-ip{\cdot}(x-y)}&
\big(\Jcal_n({p+nk})+e\frac{\ksl\asl}{4{p}{\cdot} k}\Jdn({p+nk})\big)\\&\quad\times \frac{i(\slashed{{p}}+n\ksl +m- \delta\msl)}{(p+nk)^2-m^2_*+i \epsilon}\\
&\qquad \times \big(\Jcal_n({p+nk})+e\frac{\asl\ksl}{4p{\cdot} k}\Jdn({p+nk})\big)\,.
 \end{split}
\end{align}
Note that here we have removed the bar over the mass shift matrix defined earlier in (\ref{eq:massmatrix}) since the momentum in (\ref{eq:final1}) is now off shell.
\section{Interpretation} 
\label{sec:does_this_make_sense_}
The naive expectation might have been that we would find a result of the form
\begin{equation}
\sum_{n=-\infty}^{\infty}
\int\,{\bar{}\kern-0.45em d}^{\,4}p \,e^{-ip{\cdot}(x-y)} Z^{(n)}_2\frac{i(\slashed{{p}}+n\ksl +m+\delta m)}{(p+nk)^2-m^2_*+i \epsilon}
\,,
\end{equation}
however, what we have found in (\ref{eq:final1}) looks very different. It contains matrix dependent wave function renormalisation constants on both sides of the  usual gamma matrix structure expected in a fermionic propagator as well as a minus sign in the mass shift, $\delta\msl$ in the numerator of (\ref{eq:final1}) which we will now explain.

Setting $n=0$ for simplicity and ignoring the $i \epsilon$, we make the simple observation that because the mass shift is a matrix, then it can naturally be regrouped as
\begin{equation}
	\frac1{\psl-(m+\delta\msl)}=\frac1{(\psl-\delta\msl)-m}\,,
\end{equation}
which means when we carry out the standard algebraic rearrangement we obtain
\begin{align}
\label{eq:signcaution}
\begin{split}
	\frac1{(\psl-\delta\msl)-m}&=
	\frac{\psl- \delta\msl+m}{(\psl- \delta\msl)^2-m^2}\,,\\
	&=\frac{\psl+m- \delta\msl}{p^2+e^2 a^2/2-m^2}\,,\\
	&=\frac{\psl+m- \delta\msl}{p^2-m^2_*}\,.
	\end{split}
\end{align}
Here we recognise the standard mass shift in the denominator and the minus sign present in the numerator of~(\ref{eq:final1}).

We finally define the wave function renormalisation matrix by
\begin{equation}
	\sqrt{\boldsymbol{Z}^{(n)}_2}=\Jcal_n({p+nk})+e\frac{\asl\ksl}{4p{\cdot} k}\Jdn({p+nk})\,,
\end{equation}
and note that
\begin{equation}
	\sqrt{{\overline{\boldsymbol{Z}}}^{(n)}_2}:=\gamma^0\sqrt{\boldsymbol{Z}_2^{(n)}}^{\,\dag}\gamma^0=\Jcal_n({p+nk})+e\frac{\ksl\asl}{4p{\cdot} k}\Jdn({p+nk})\,.
\end{equation}
This yields a very appealing way to write the sideband  propagator. If we  
reinstate the sum over sidebands, we see that the Volkov propagator for fermions can be written as
\begin{align}\label{eq:final}
\begin{split}
	\sum_{n=-\infty}^\infty \braV{0}\mathrm{T}\psi_n(x)\bar{\psi}_n(y)\ketV{0}
=\sum_{n=-\infty}^\infty\int\,{\bar{}\kern-0.45em d}^{\,4}p \,& e^{-ip{\cdot}(x-y)}\sqrt{\overline{\boldsymbol{Z}}^{(n)}_2}\frac{i}{\psl+n\ksl -(m+\delta\msl)+i \epsilon}\sqrt{{{\boldsymbol{Z}}}^{(n)}_2}\,.
\end{split}
\end{align}

\section{Does this agree with perturbation theory?} 
\label{sec:does_this_agree}
We will now check that our result for the propagator (\ref{eq:final}) agrees with perturbation theory by calculating the lowest order corrections to the free propagator extending our earlier result~(\ref{eq:pertcentraltwo}) to the adjacent sidebands. 

We will require the expansions of the generalised Bessel functions 
\begin{align}
\begin{split}
\Jcal_{0}(p)=\mathrm{J}_0(eu_p,e^2v_p)&=1-\frac{e^2u_p^2}{4}+{\cal{O}}(e^4)\,,  \\
\Jcal_{\pm 1}(p)=\mathrm{J}_{\pm1}(eu_p,e^2v_p)&=\pm\frac{eu_p}{2}+{\cal{O}}(e^3)
\,.
\end{split}
\end{align}
We should immediately  note that (\ref{eq:final}) is indeed an expansion in $e^2$. In~(\ref{eq:final1}) we see that for any value of $n$ the  bracketed terms containing the combinations of generalised Bessel functions are either both odd or both even in $e$ due to the property $\mathrm{J}_n(-x,y)=(-1)^n\mathrm{J}_n(x,y)$. As there are two bracketed terms,  the overall propagator will  be an expansion in $e^2$.

Let us first consider the central sideband at leading order. The wave function renormalisation factors are
\begin{equation}
\sqrt{{{\boldsymbol{Z}}}^{(0)}_2}=	\sqrt{{\overline{\boldsymbol{Z}}}^{(0)}_2}
 =
	1-e^2\left(\frac{p{\cdot}a}{p{\cdot}k}\right)^2,
\end{equation}
as the matrix factors cancel at this order here. Using $\delta\msl$  it is now simple to reproduce all of~(\ref{eq:pertcentraltwo}) from an expansion of~(\ref{eq:final}). This calculation is sensitive to both the matrix form of the mass shift and the sign in the numerator of~(\ref{eq:signcaution}).

In the first upper sideband we proceed similarly to~(\ref{eq:gammatrick}) but adding and subtracting $\ksl$~terms as follows
\begin{align}
	(\psl+m)\asl(\psl+\ksl+m)\asl(\psl+m)=
(\psl+m)\asl(\psl+\ksl+m)\asl(\psl+\ksl+m-\ksl)
\,,
\end{align}
which may also be repeated on the left. This  generates zero pole terms that cancel with those found in the central sideband and from the first lower sideband. The terms containing poles in the first upper sideband are thus found to be
\begin{align}\label{eq:pertuppertwo}
 \begin{split}
	ie^2 \Bigg(
\frac14\left(\frac{p{\cdot} a}{p{\cdot} k}\right)^2 & \frac{\psl+\ksl+m}{(p+k)^2-m^2}
+
\frac14\frac{p{\cdot} a}{p{\cdot} k}\asl\frac{1}{(p+k)^2-m^2} \\
& -
\frac14\left(\frac{p{\cdot} a}{p{\cdot} k}\right)^2 \ksl\frac{1}{(p+k)^2-m^2}
-
\frac18\frac{p{\cdot} a}{p{\cdot} k} a^2 \ksl
\frac{1}{(p+k)^2-m^2}
\Bigg)\,.
 \end{split}
\end{align}
 and similarly with $k\to-k$ for the first lower sideband. (The adjacent sidebands will first acquire mass shifts at the next order in the coupling.)

This calculation also agrees with~(\ref{eq:final}) at this order. In the first upper  sideband, we note that 
\begin{equation}\label{eq:ztwonisone}
	\sqrt{{\overline{\boldsymbol{Z}}}^{(1)}_2}=
	e\left(-\frac12\frac{p{\cdot}a}{p{\cdot}k}+
\frac{\ksl\asl}{4p{\cdot}k}
\right)
\,,
\end{equation}
at leading order in $e$. There is a similar expression for $\sqrt{{{\boldsymbol{Z}}}^{(1)}_2}$. Substituting this into~(\ref{eq:final}) yields~(\ref{eq:pertuppertwo}). The final $a^2$-dependent term in~(\ref{eq:pertuppertwo}) may appear surprising in this sideband as there is no contribution from the mass shift in the perturbative expansion at this order. It  arises via a product of two $\asl$-factors in the gamma matrix structures.    We note that this calculation is sensitive to the matrix structure of the wave function renormalisation factors. The first lower sideband is again simply reproduced by changing $k\to-k$ everywhere     in~(\ref{eq:pertuppertwo}).

As the agreement across these three sidebands depends upon the form of the various structures in~(\ref{eq:final}), it verifies the emergence of a matrix mass shift and matrix wave function structures ordered to the left and right. 

We also comment here that the propagator in the laser background cannot be written in the form
\begin{equation}\label{eq:selfenergy}
	\frac i{\psl-m-\Sigma}\,,
\end{equation}
with $-i\Sigma$ being the self-energy. This is because the usual derivation of~(\ref{eq:selfenergy}) involves the identification of the perturbative expansion 
\begin{equation}
	\frac i{\psl-m}+\frac i{\psl-m}(-i\Sigma)
	\frac i{\psl-m}+\cdots\,,
\end{equation}
as a Taylor expansion of~(\ref{eq:selfenergy}). However, this  requires that $\psl$ and $\Sigma$ commute. This is true in normal perturbative QED since the self-energy then only depends on $\psl$ and scalars. However, in the laser context, there are $\ksl$ and $\asl$ factors which do not commute with $\psl$ and hence obstruct an expression of the form~(\ref{eq:selfenergy}).

\section{Conclusions} 
\label{sec:conclusions}
In this paper we have derived an expression for the fermion propagator in a linearly polarised plane wave. In QED, as well as in the scalar theory, all of the laser induced structures in the propagator can be understood as multiplicative renormalisations of the free propagator. 
The distinctive feature highlighted in this paper is that for fermionic QED these multiplicative factors are matrices. Matrix structures have already been seen in earlier work~\cite{Reiss:1966A}\cite{Brown:1964zz}\cite{Ritus:1985review}, but here we have seen that  the electron mass shift is a matrix such that both the mass and also the mass squared only pick up corrections at leading order in the coupling.  Although we have focused in this paper on linearly polarised lasers, our approach can be generalised to other polarisations, see~\cite{Lavelle:2013wx}. 
We expect that such a matrix valued mass shift will persist in more realistic models of laser pulses~\cite{Harvey:2012ie}.  
We note that matrix wave function renormalisation constants have been seen previously in studies of the infra-red~\cite{Bagan:1999jk} although the mass shift did not there involve a matrix.  We have also checked our all orders calculations through a first order perturbative expansion of the propagator which tested all of the key features of our overall result~(\ref{eq:final}).

The results presented in this paper will be important for calculating scattering processes in QED in laser backgrounds, see for example~\cite{King:2014wfa} where the propagator is used to calculate double Compton scattering,  and also~\cite{Seipt:2012tn}\cite{Mackenroth:2012rb} for this process. More generally see~\cite{Heinzl:2010vg}\cite{Ilderton:2010wr}\cite{hu10}\cite{king13b} for other processes. It is intriguing to speculate how the matrix renormalisation constants impact upon the calculation of S-matrix elements and the physical predictions of QED in a laser background.


\appendix

\section{Volkov Spinors} 
\label{sec:volkov_spinors}
The defining equations~(\ref{eq:uvolkov}) and~(\ref{eq:vvolkov}) allow us to construct the Volkov spinors by suitably boosting the static spinors
\begin{equation}
		\U^{(1)}(0)=\begin{pmatrix}
			1\\0\\0\\0
		\end{pmatrix},\quad
		\U^{(2)}(0)=\begin{pmatrix}
			0\\1\\0\\0
		\end{pmatrix},\quad
		\V^{(1)}(0)=\begin{pmatrix}
			0\\0\\1\\0
		\end{pmatrix},\quad
		\V^{(2)}(0)=\begin{pmatrix}
			0\\0\\0\\1
		\end{pmatrix}.
	\end{equation}
To this end,  we define 
\begin{equation}
	\Uv^{(\alpha)}(p)=\frac1{\sqrt{2m(m+E^*_p+2e^2v_\pbar k_0)}}(\pbsl+2e^2v_\pbar\ksl+m)\U^{(\alpha)}(0)\,,
\end{equation}
and
\begin{equation}
	\Vv^{(\alpha)}(p)=\frac1{\sqrt{2m(m+E^*_p+2e^2v_\pbar k_0)}}(-\pbsl-2e^2v_\pbar\ksl+m)\V^{(\alpha)}(0)\,.
\end{equation}
With these conventions we  have the inner products 
\begin{equation}
	\Uvb^{(\alpha)}(p)\Uv^{(\beta)}(p)=\delta^{\alpha \beta}\,,\quad\mbox{and}\quad
	\Vvb^{(\alpha)}(p)\Vv^{(\beta)}(p)=-\delta^{\alpha \beta}\,,
\end{equation}
as well as the tensor products
\begin{equation}
	\Uv^{(\alpha)}(p)\Uvb^{(\alpha)}(p)=\frac{\pbsl +2e^2v_\pbar\ksl+m}{2m}\,,
\end{equation}
and 
\begin{equation}
	\Vv^{(\alpha)}(p)\Vvb^{(\alpha)}(p)=\frac{\pbsl +2e^2v_\pbar\ksl-m}{2m}\,.
\end{equation}

\section{Volkov propagator details} 
\label{sec:volkov_propagator_details}
The argument taking us from the initial expression for the Volkov propagator in Equation~(\ref{eq:vstep1}) to the final covariant expression (\ref{eq:final1}) involves several key steps that we outline below.

Starting from~(\ref{eq:vstep1}) we make the substitution $p^0\to E^*_p-p^0-nk^0$ in the first part and $p^0\to p^0+E^*_p+nk^0$ in the second to give
\begin{align}
\begin{split}
-\int\!\frac{{\bar{}\kern-0.45em d}^{\,4}p}{2E^*_p}&\Big(\frac{i}{E^*_p-p^0-nk^0-i \epsilon}
e^{-ip^0(x^0-y^0)}e^{i(\underline{p}-n\underline{k})\cdot(\underline{x}-\underline{y})}\\
&\times \big(\Jcal_n(\pbar)+e\frac{\ksl\asl}{4\pbar{\cdot} k}\Jdn(\pbar)\big)(\pbsl+m- \delta\mslb)\big(\Jcal_n(\pbar)+e\frac{\asl\ksl}{4\pbar{\cdot} k}\Jdn(\pbar)\big)\\
&\qquad -\frac{i}{E^*_p+p^0+nk^0-i \epsilon}e^{-ip^0(x^0-y^0)}e^{-i(\underline{p}+n\underline{k})\cdot(\underline{x}-\underline{y})}\\
&\times \big(\Jcal_n(-\pbar)+e\frac{\asl\ksl}{4\pbar{\cdot} k}\Jdn(-\pbar)\big)(\pbsl-m- \delta\mslb)\big(\Jcal_n(-\pbar)+e\frac{\ksl\asl}{4\pbar{\cdot} k}\Jdn(-\pbar)\big)
\Big)\,.
\end{split}
\end{align}
If we now shift the three momentum in the first part by $\underline{p}\to \underline{p}+n\underline{k}$ and by $\underline{p}\to -\underline{p}-n\underline{k}$ in the second (noting a change in sign in $\delta\mslb$) then we have
\begin{align}
\begin{split}
-\int\!\frac{{\bar{}\kern-0.45em d}^{\,4}p}{2E^*_{p+nk}}e^{-ip{\cdot}(x-y)}&\Big(\frac{i}{E^*_{p+nk}-p^0-nk^0-i \epsilon}
\big(\Jcal_n(\overline{p+nk})+e\frac{\ksl\asl}{4\overline{(p+nk)}{\cdot} k}\Jdn(\overline{p+nk})\big)\\&\qquad\quad({\overline{\psl+n\ksl}} +m- \delta\mslb)\big(\Jcal_n(\overline{p+nk})+e\frac{\asl\ksl}{4\overline{(p+nk)}{\cdot} k}\Jdn(\overline{p+nk})\big)\\
&-\frac{i}{E^*_{p+nk}+p^0+nk^0-i \epsilon}\big(\Jcal_n(\widetilde{p+nk})+e\frac{\asl\ksl}{4\widetilde{(p+nk)}{\cdot} k}\Jdn(\widetilde{p+nk})\big)\\&\qquad\quad(-\widetilde{\psl+n\ksl}-m+\delta\mslt)\big(\Jcal_n(\widetilde{p+nk})+e\frac{\ksl\asl}{4\widetilde{(p+nk)}{\cdot} k}\Jdn(\widetilde{p+nk})\big)
\Big)\,.
\end{split}
\end{align}
In this we have simply shifted on-shell momenta in the first part but in the second part noted that under the shift
\begin{align}
\begin{split}
	\pbar&\to (E_{p+nk},-\underline{p}-n\underline{k})\,,\\
	&=-(-E_{p+nk},\underline{p}+n\underline{k})\,,\\
	&=-\widetilde{p+nk}\,,
	\end{split}
\end{align}
where we denote by the tilde the negative energy on-shell four-momenta.

Combining these terms over a common denominator allows us to extract the covariant propagator and we get
\begin{align}
\begin{split}
\int \!\frac{{\bar{}\kern-0.45em d}^{\,4}p}{2E^*_{p+nk}}&\frac{ie^{-ip{\cdot}(x-y)}}{(p+nk)^2-m^2_*+i \epsilon}\Big((E^*_{p+nk}+p^0+nk^0)
\big(\Jcal_n(\overline{p+nk})+e\frac{\ksl\asl}{4\overline{(p+nk)}{\cdot} k}\Jdn(\overline{p+nk})\big)\\&\qquad\quad({\overline{\psl+n\ksl}} +m- \delta\mslb)\big(\Jcal_n(\overline{p+nk})+e\frac{\asl\ksl}{4\overline{(p+nk)}{\cdot} k}\Jdn(\overline{p+nk})\big)\\
&-(E^*_{p+nk}-p^0-nk^0)\big(\Jcal_n(\widetilde{p+nk})+e\frac{\asl\ksl}{4\widetilde{(p+nk)}{\cdot} k}\Jdn(\widetilde{p+nk})\big)\\&\qquad\quad(-\widetilde{\psl+n\ksl}-m+ \delta\mslt)\big(\Jcal_n(\widetilde{p+nk})+e\frac{\ksl\asl}{4\widetilde{(p+nk)}{\cdot} k}\Jdn(\widetilde{p+nk})\big)
\Big)\,. 
\end{split}
\end{align} 
We now recognise the two terms as simply the residues associated with the two poles of the covariant denominator. Hence we recover the covariant expression~(\ref{eq:final1}).
 
\bigskip

\noindent \textbf{Acknowledgements:} We thank Tom Heinzl, Anton Ilderton and Ben King for helpful discussions.

\newpage


%

\end{document}